\documentclass[Royal,sagev,times]{sagej}

\usepackage{moreverb,url}

\usepackage[colorlinks,bookmarksopen,bookmarksnumbered,linkcolor=black,citecolor=black,urlcolor=blue]{hyperref}

\usepackage[utf8]{inputenc}
\usepackage[english, german]{babel}
\usepackage{rotating,tabularx}
\usepackage{longtable}
\usepackage{pdflscape}
\usepackage{supertabular}

\def\degr{\hbox{$^\circ$}}
\def\arcmin{\hbox{$^\prime$}}
\def\arcsec{\hbox{$^{\prime\prime}$}}

\def\volumeyear{2017}

\begin{document}
\selectlanguage{english}

\runninghead{Remchin and Schrimpf}

\title{Research on asteroids of Christian Ludwig Gerling and his students in the
nineteenth century}

\author{Julia Remchin\\
{\footnotesize \normalfont{Independent Scholar, Germany}}\\
\vspace{1em}
Andreas Schrimpf\\
{\footnotesize  \normalfont{Philipps--Universität Marburg, Germany}}
}


\corrauth{Andreas Schrimpf, 
History of Astronomy and Observational Astronomy,
Philipps-Universität Marburg,
Renthof 7b,
D-35032 Marburg,
Germany.}

\email{andreas.schrimpf@physik.uni-marburg.de}

\begin{abstract}
One of the mayor topics in astronomy at the beginning of the 19th century was the interpretation of the observations of the first asteroids. In 1810 Christian Ludwig Gerling at the age of twenty two came to Göttingen University to continue his academic studies. Supervised by Carl Friedrich Gauß at the observatory he was engaged in studies of theoretical and practical astronomy.  
Starting in 1812 Gerling 
 accepted the responsibility for collecting observational data of the asteroid Vesta from the European observatories and for calculating the ephemeris of this new minor planet. In 1817 Gerling was appointed professor at Marburg University. One of his early astronomical projects in Marburg was his contribution to the \emph{Berliner Akademische Sternkarten}.
After completion of his observatory in 1841 Gerling's students started observing and theoretically analysing the orbits of the continuously newly discovered asteroids including the perturbation of the larger solar system bodies. The observations at Gerling’s observatory are the first astrometric measurements of solar systems minor bodies of Hesse.
\end{abstract}

\keywords{Asteroids, star maps, orbit calculation of asteroids, astronomy in the nineteenth century, Gerling-Observatory, Marburg}

\maketitle

\section{Introduction}

The starry sky was and still is fascinating. Since ages all cultures have been studying
the motions of stars and solar system bodies. At the edge of modern ages based on observations without telescopes the positions and motions of Sun, Moon and the planets could be predicted with considerable precision. Kepler's laws and the development of telescopes increased the knowledge of the orbits of solar system bodies to a great extend leading to a new view of world, the Copernican system, which finally became accepted after the publication of Newton's law of gravity. 

The discovery of Uranus, the seventh planet, in the year 1781 by Wilhelm Herschel triggered  an euphoric search for further solar system bodies. In fact, the first four asteroids were discovered at the beginning of the nineteenth century between 1801 and 1807 and the story of their discovery was studied and told by many authors from the very beginning, see for example G. Jahn \cite{Jahn1844}, to a very recent and very detailed exploration by C. Cunningham. \cite{Cunningham2017}

The orbits of the first four asteroids roughly are located at the same region between Mars and Jupiter. The sizes of these asteroids were not known at that time so they were reckoned as ordinary planets. But this clearly opened a new problem: the Titius--Bode law\cite{Bode1802} predicted one planet between Mars and Jupiter not four. Were the asteroids fragments of a former collision? To answer these questions further observations  seemed to be necessary. All observatories in Europe monitored the new four planets. Besides Carl Friedrich Gauß his students Friedrich Bernhard Gottfried Nicolai, Johann Franz Encke, Friedrich Ludwig Wachter, Christian Ludwig Gerling and August Ferdinand Möbius calculated new orbital elements and ephemerides. Wachter and Möbius took care of Juno, Encke was responsible for Pallas and Gerling was busy with Vesta.\cite{Jahn1844b1}

It was clear, further solar system bodies had to be less bright than the ones known, otherwise they would have been detected till then. Therefore one of the major demands was the improvement of star charts, i.e. less bright stars should be included. The catalogues of Jérôme Lalandes, the  Histoire Céleste, and Giuseppe Piazzi increased the number of recorded stars to about 50,000.\cite{Encke1826a} However, those catalogues were not complete, they did not contain all stars down to a certain brightness. Hence in 1824 Friedrich Wilhelm Bessel initiated the compilation of the \emph{Berliner Akademische Sternkarten}\cite{Bessel1824}, thus supporting the search for new smaller solar system bodies.
 
For many years following 1807 no new minor planet was discovered. Then, in December 1845, 38 years after the first observation of Vesta, Karl Ludwig Hencke discovered the fifth minor planet Astraea. Though the new solar system bodies are much smaller than any other planet, they had been counted as regular planets. When Neptune was discovered in 1846 it became the 13th planet of our solar system.\cite{Humboldt1850} In July 1847 the sixth minor planet was discovered, and then no year passed without new discoveries. Accepting the proposal of Alexander von Humnboldt\cite{Humboldt1850b1} in 1851 a new class of solar system bodies, finally was introduced. The minor planets were named \emph{asteroids}, just as Herschel had suggested already in 1802. 

With the increasing number of discovered asteroids and the increasing precision of observations the perturbations of the large solar system bodies on the orbits of the asteroids came into focus, which finally lead to the discovery of the Kirkwood gaps and the chaotic properties of asteroids in the main belt.\cite{Murdin2016}

In this paper we will review the up to day somewhat overseen contribution of
Christian Ludwig Gerling in the beginning of the research on asteroids. His ephemeris calculations have been shortly summarized by G. Jahn \cite{Jahn1844b2} and were cited by C. Cunningham.\cite{Cunningham2017_2} However his contributions to the improvement of the star maps and his observatory at Marburg and the observations of his students in the mid of the 19th century as well as the theoretical investigations of perturbations of the larger planets on the orbits of smaller asteroids were not recognized so far. 
The paper presented
here is based on publicly available information and the scientific estate of Chr. L. Gerling containing his correspondence and notes.\cite{Gerling_archive}

\section{Christian Ludwig Gerling} 
\begin{figure}
	\centering
	\includegraphics[width=0.6\linewidth]{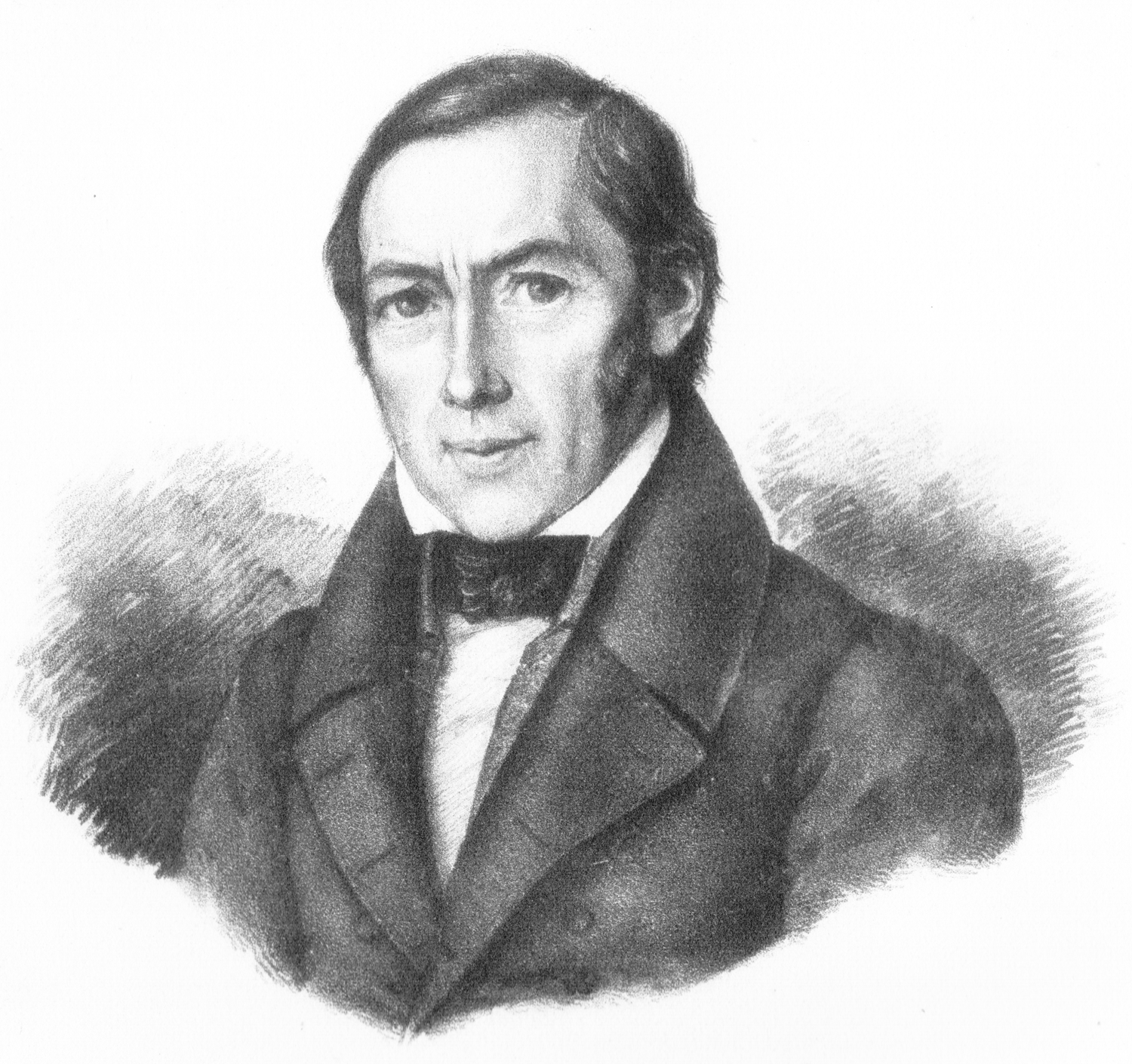}
	\caption{Christian Ludwig Gerling (1788-1864)}
	\label{fig_gerling}
\end{figure}

Christian Ludwig Gerling (Fig.\ \ref{fig_gerling}) was born in Hamburg, Germany, in 1788 as a son of a preacher. His education started with private lessons at home and continued at the grammar school ``Johanneum'', Hamburg, together with his long--time friend Johann Franz Encke. After school Gerling registered at the University of Helmstedt, Lower Saxony, for theology. In addition he attended lectures in mathematics of professor Pfaff. In 1810 the University of Helmstedt was closed and Gerling decided to continue his academic studies in mathematics and astronomy in Göttingen. At the observatory he kept himself busy with theoretical and practical astronomy under supervision of Carl Friedrich Gauß and Karl Ludwig Harding.\cite{Bruns1879} There he got to know Gottfried Nicolai, who later became director of the observatory in Mannheim, Germany, and Friedrich Ludwig Wachter. Encke joined this group in 1811. He later became director of Seeberg Observatory, Gotha, and finally in 1825 of the Berlin Observatory.

Studying under Gauß Gerling not only learned the appropriate use of instruments but also the mathematical methods for analysis of geodetic and astronomic measurements. Later, Gerling published among others a book about the practical use of least squares in geodetic data analysis.\cite{Gerling1843} 

In 1812 Gerling finished his doctoral thesis about a calculation of the path of the solar eclipse in 1820 in northern Europe. After he received his PhD 
Gerling entered a position at a high school in Kassel,
Hesse. At that time he used a small observatory in Kassel for astronomical
observations. He continued to seek a university position and finally in
1817 was appointed full professor of mathematics, physics and astronomy and director of the ''Mathematisch--Physikalisches Institut'' at the Philipps--Universität of Marburg.
In spite of several offers he remained at the university in Marburg till his
death in 1864.\cite{Madelung1996} During his career three times he kept the position as prorector of the university, he was representative in the chamber of the Electorate of Hesse and in 1857 he was appointed court counsellor. Due to his scientific reputation he was nominated member of several scientific societies, e.g. the Göttingen Royal Society of Sciences.\cite{Bruns1879b1}

Gerling's scientific work in Marburg was affected by two major topics: in his early
period from 1817 to 1838 he was well occupied with organizing
the triangulation of Kurhessen.\cite{Reinhertz1901} During that time Gerling did not have an
observatory, so his astronomical work was reduced mainly to corrections and additions
on a section of Encke's ``Berliner Akademische Sternkarten''. 
In 1838 the institute moved to a new home in Marburg at the
''Rent\-hof''. After the building was reconstructed in 1841, he
could finally put into operation his new but small observatory, built on top of
a tower of Marburg's old city wall.\cite{Schrimpf2010}
Gerling pursued the scientific topics of astronomy of that time, making
meridional observations and differential extra--meridional measurements of stars, planets and asteroids, observations of lunar occultations, as well as astronomical time determinations mainly to improve the precision of star catalogues, improving orbital parameters of solar system bodies and the solar parallax.

Carl Friedrich Gauß and Christian Ludwig Gerling started their relationship as a teacher and his student but during the following years they became each others counsellor and finally close friends. Their correspondence not only contains details of scientific discussions but also reflects their close relationship.\cite{GGLetters} 
Gerling infected his students with his enthusiasm for astronomy: Carl Wilhelm Moesta became the first director of the National Observatory of Chile,\cite{Schrimpf2014} Eduard Schönfeld became director of the observatory in Mannheim, Germany, and later the successor of Friedrich Wilhelm August Argelander in Bonn, and Ernst Wilhelm Klinkerfues became assistant and later successor of C.F. Gauß in Göttingen.
Gerling died in 1864 at the age of 76. His estate has been
stored in the library of the Philipps--Universität Marburg.
About 1000 letters of his correspondence with Gauß, Encke, Nicolai and many more give an insight into his personal and professional life
and can be a valuable source for historians.\cite{Gerling_archive_2}

\section{Gerling's Vesta calculations}

Gerling calculated his first Vesta ephemeris for 16 Juli 1812 to 30 April 1813 in 1811, while he still was in Göttingen with C.F. Gauß. It was published in the \emph{Monatliche Correspondenz}.\cite{Gerling1811} After starting his job in Kassel Gerling struggled to get access to a small observatory, which was not in good shape. In October 1813 he observed Vesta at RA = 133\degr, DEC = 18\degr20\arcmin\  and marked it in Harding's star map, not really being convinced that the object he had observed 
was Vesta.\cite{Gerling_28} 
On 17 February 1814 he send a note to Gauß, that to his surprise he observed Vesta just at the position his own ephemeris had predicted.\cite{Gerling_39} 

\begin{figure}
	\centering
	\includegraphics[width=0.8\linewidth]{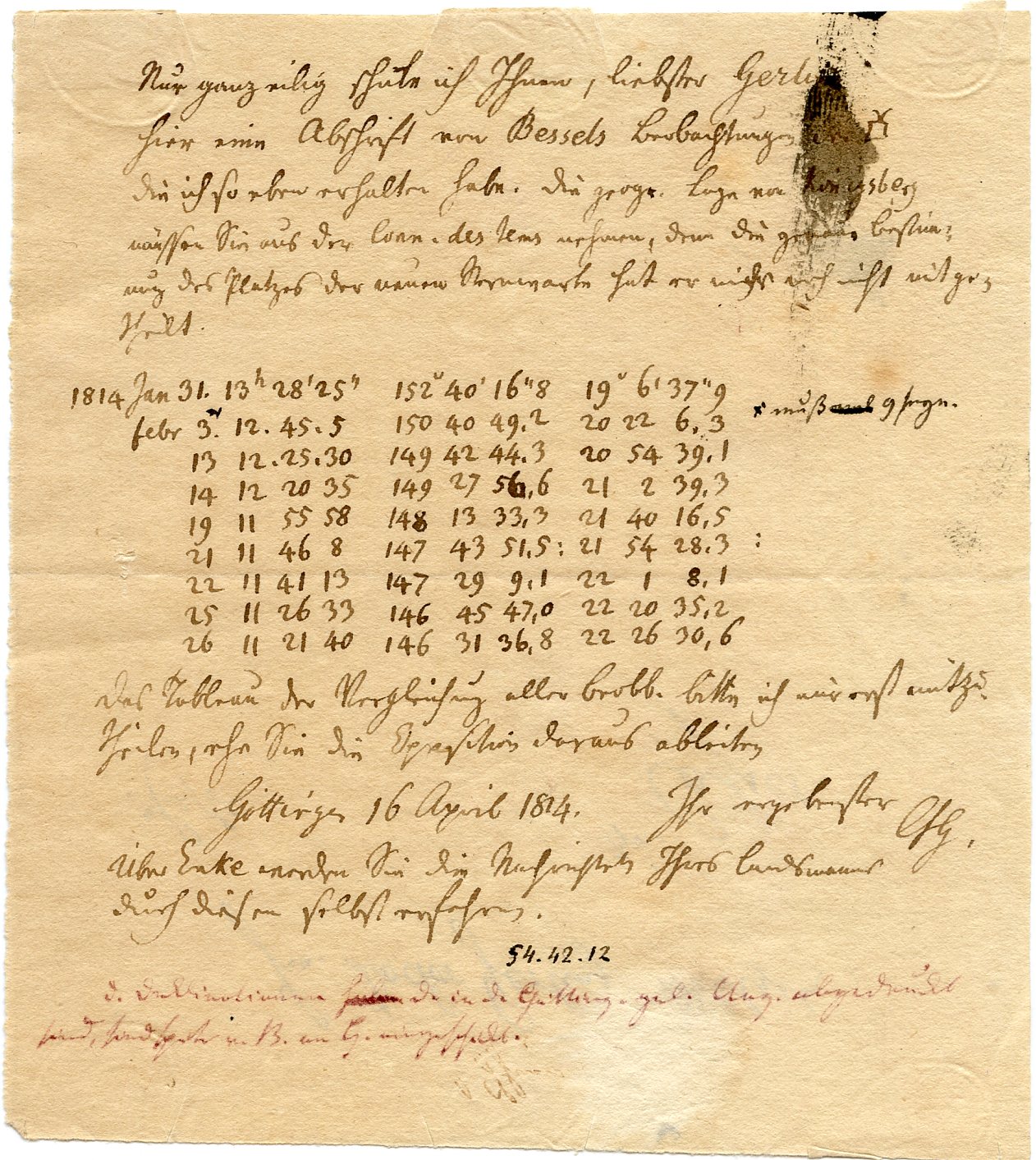}
	\caption{Letter from C.F. Gauß to Gerling dated 16 April 1814. The letter contains the Vesta observational data from F.W. Bessel\cite{Gauss1814}, see transliteration in the appendix.}
	\label{fig_letter_gauss_1814}
\end{figure}

In the same letter Gerling asked Gauß for observational data of Vesta in order to take care of further orbital calculations of this asteroid, if Gauß had not appointed this task already to someone else. Gauß appreciated Gerling's offer and sending his own observation taken with a mural instrument as well as observations from Nicolai and Bessel (Fig.\ \ref{fig_letter_gauss_1814}) he asked Gerling to calculate the orbital elements of Vesta based on the four oppositions from 1810 to 1814.\cite{Vesta_10_14} 
Gerling's results were published in the \emph{Göttinger Gelehrte Anzeigen}, the journal of the Göttingen Royal Society of Sciences, and in Bode's \emph{Berliner Astronomisches Jahrbuch},\cite{Gerling1814a} where Gerling added a forecast of the oncoming Vesta opposition in July 1815. By the end of August 1814 Gerling finished the calculation of 
the ephemeris of Vesta in 1815 and sent them to Gauß, Harding and Bode, just in time to be included in the \emph{Berliner Astronomisches Jahrbuch}.\cite{Gerling1814b} Gerling commented his results in his letter to Gauß, stating that "due to the small distance of Vesta in the next opposition its brightness would be extraordinary if it would not have such a large southern declination".\cite{Gerling_51} 

In summer 1815 Gerling received observational data of that Vesta opposition from Seeberg Observatory by  Bernhard v.\ Lindenau and Nicolai, who at that time was working at Seeberg Observatory, as well as from Königsberg Observatory by Bessel, from Mannheim Observatory by Friedrich Georg Wilhelm Struve, visiting at Mannheim, and from Milan Observatory by Francesco Carlini.\cite{Vesta_int_15} 
The final 1815 oppositional data were published in the \emph{Berliner Astronomisches Jahrbuch}.\cite{Gerling1816a}

Vesta reached its sixth opposition on 4 December 1816. In fall 1815 Nicolai\cite{Nicolai1815} sent Bode's Vesta observations to Gerling and by the end of March 1816 Gerling finished the new ephemeris for August 1816 to March 1817 and sent them to Gauß\cite{Gerling_82}
and Bode.\cite{Bode1816} 
In the new volume of the \emph{Berliner Astronomisches Jahrbuch} ephemeris, orbital elements, dates for eastern and western quadrature and the opposition, as well as the brightness for all oppositions from 1808 to 1816 were published\cite{Gerling1816b}. Again Gerling received observational data of the Vesta opposition from many European observatories (Mannheim, Seeberg, Königsberg, Berlin, Wien, Milan),\cite{Vesta_int_16} 
which he used to calculate the final data for the opposition on 4 December 1816.\cite{Gerling1817a}

In May 1817 Gerling was appointed full professor at Marburg University. His new tasks took most of his time. In fall 1817 he calculated his last Vesta ephemerides for dates from 2 January 1818 to 21 July 1818 and sent them to Bode for publishing\cite{Gerling1817c} and to Gauß.\cite{Gerling_106}
Encke continued the Vesta calculations in the following years.

Orbital calculations obviously were very time consuming in the 19th century. More than once Gauß was faced with cancellations by his colleagues. In 1816 he was in need for someone to follow Juno. He sent Gerling a few observations of Juno from Harding and Gerling calculated the opposition.\cite{Juno16}
However, Gerling asked Gauß to pass this task to someone else and suggested Encke, but finally Nicolai continued the Juno calculations for several years.

During the years Gerling assumed the responsibility for the Vesta calculations he was in  continuous correspondence with Gauß, Encke, Nicolai, Bode and v.\ Lindenau. The letters he received are stored in the archive of Christian Ludwig Gerling in the library of the Philipps-University Marburg, one of them is shown in Fig. \ref{fig_letter_gauss_1814}. Some of the results of Gerling's calculations were also published in \emph{Zeitschrift für Astronopmie und verwandte Wissenschaften} edited by B.\ von Lindenau and J.\ G.\ F.\ Bohnenberger.

\section{Gerling's contribution to the \emph{Berliner Akademische Sternkarten}}

After the discovery of the first four asteroids it soon became obvious, that further less bright objects could be found more easily if the star charts would contain more less bright stars. In 1824 Bessel initiated the compilation of the \emph{Berliner Akademische Sternkarten},\cite{Bessel1824b1} a belt from 15\degr\ north to 15\degr\ south of the equator containing all the stars down to a brightness of 9 mag to 10 mag. On 1 November 1825 Encke published a call in the \emph{Astronomische Nachrichten} to join that campaign. Every participating observatory or astronomer should take care of a chart 15 degree (1 hour) wide and was promised a reward of 25 Dutch guilders.\cite{Encke1826ab1} Gerling offered his help and the committee of the academy agreed that Gerling should take care of the star field from RA 15:56 to RA 17:04, i.e. Hora 16.\cite{Encke1826b}

At that time Gerling did not have an observatory with a larger telescope at his disposal. However, this was not urgently necessary for the improvement of the star charts, In November 1826 Gerling sent a note to Gauß that he had built himself an equatorial mount for his smaller telescope\cite{Gerling_174} 
to participate in that project. Two years later in November 1828 he told Gauß that his observations for this campaign had taken all his time in summer 1828.\cite{Gerling_178} 
However, Gerling's activity must have been stalled. In 1831 he received a letter from Encke on behalf of the committee urging him to continue his part of the work. In the interim report dated 1833 Encke mentioned that Gerling still was willing to finish his part of the chart but he could not give a final date.\cite{Encke1834} Gerling was not able to resume his contribution and in 1840 he finally ceased his participation in the star charts. Encke passed Gerling's Hora 16 to Dr. Jakob Philipp Wolfer, Berlin, who finished this part in 1843. 
In his epilogue Wolfers thanked Gerling sincerely,\cite{Wolfers1843}
\begin{itemize}
\item 
Professor Gerling of Marburg had begun to work on the present chart before I did but was prevented from completing it. When he learned I had taken over the process, he was more than willing to place all the work he had accomplished thus far at my disposal and this proved to be an essential service particularly in the reduction of stars. In some ambiguous cases I was able to compare with Gerling’s reduction and readily attain certainty. I therefore feel compelled at this occasion to express my most profound gratitude to Professor Gerling. 
\end{itemize}

\section{The Gerling-Observatory at Marburg University}
\label{sec_gerling_observatory}

When Gerling accepted his professorship at Marburg University in 1817 the physical and astronomical section were not in an appropriate organized state. A well constructed institute including an observatory was of particular concern to Gerling. He had to wait until 1838 before the university offered him the \emph{Dörnberger Hof} at the castle hill, a former estate for supply of the landgraves in the near castle. The massive reconstruction including the provision of lecture hall and laboratories for the institute, of the apartment for the institute's director and of an observatory on top of the tower lasted until 1841.\cite{Gerling_317} 

\begin{figure}
	\centering
	\includegraphics[width=1.0\linewidth]{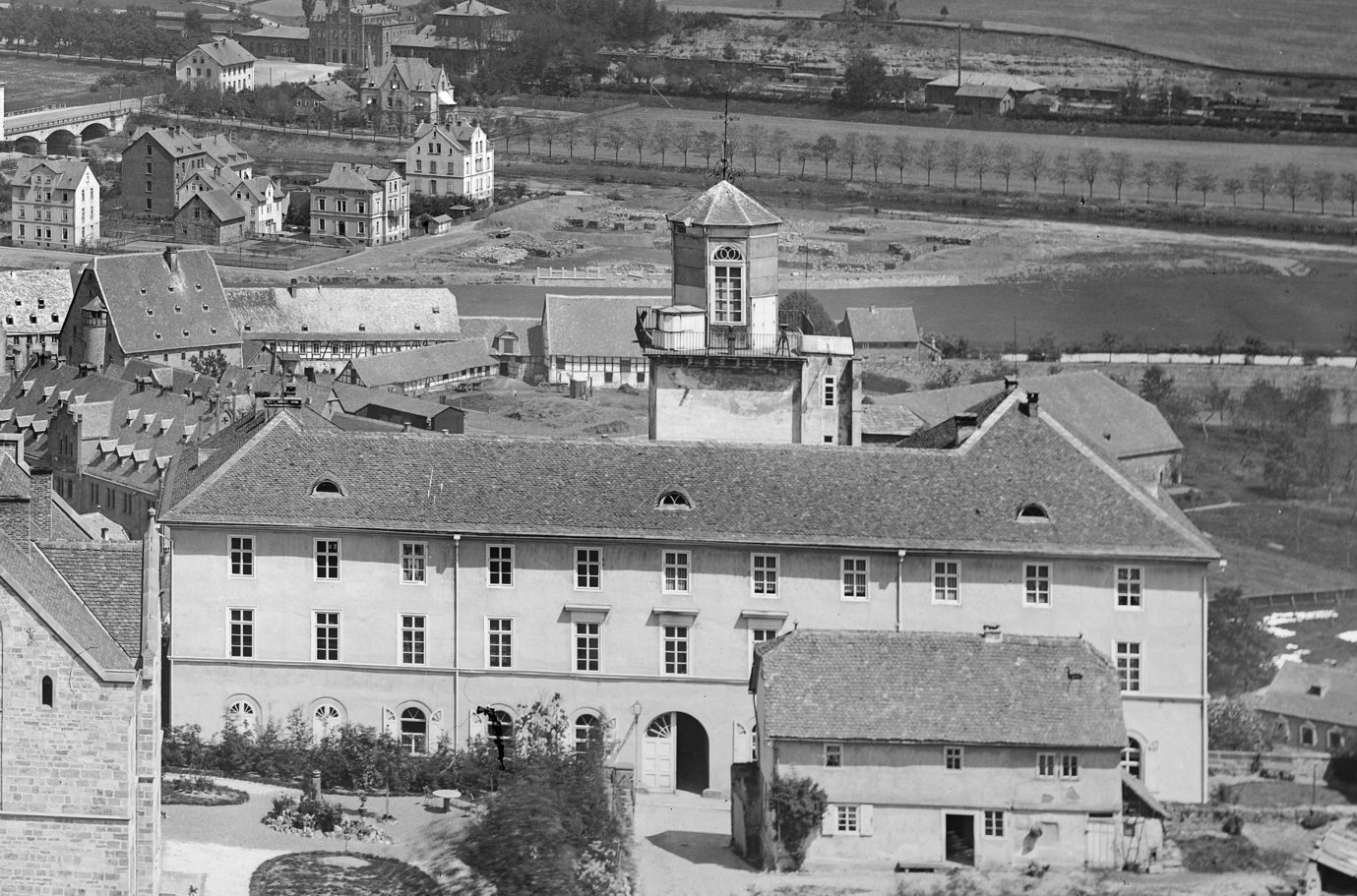}
	\caption{The \emph{Mathematisch--Physikalische Institut} at Renthof, 1880. The observatory is located on top of the tower in the back of the building. (Photo: Bildarchiv Foto Marburg. Photographer: Ludwig Bickell)}
	\label{fig_gerling_observatory}
\end{figure}

In the upper part of the tower Gerling established his observatory, consisting of an octagonal small house surrounded by a gallery (Fig.\ \ref*{fig_gerling_observatory}). A post was mounted at the gallery. It was supported by a vault underneath, to provide a solid and low vibrating mount for the telescope. To protect the telescope a partly removable case was constructed, easily visible on the left in front of the observatory house  (Fig.\ \ref*{fig_gerling_observatory}). Unfortunately this case later was demounted, it does not exist any more. From this post Gerling and his students could observe the southern, western and northern parts of the sky. For observations in the prime vertical (i.e. East and West) the windows of the small house were opened to the East. Thus observations up to a zenith distance of 12\degr\ in eastern direction could be made.

\begin{figure}
	\centering
	\includegraphics[width=0.6\linewidth]{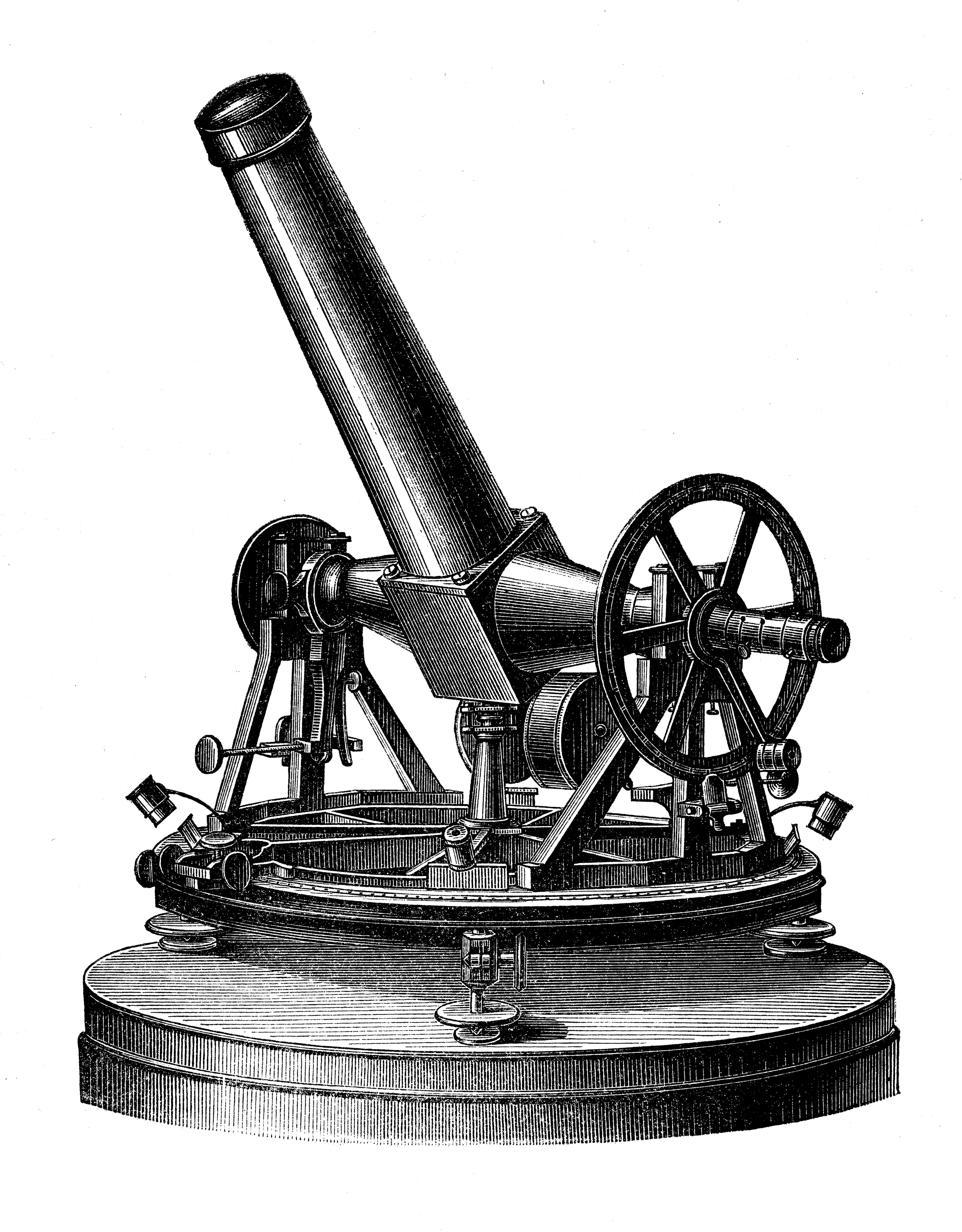}
	\caption{Gerling's transit telescope, Ertel \& Sohn,  1841. Reproduction from \emph{Astronomische Zeitbestimmungen} \cite{Melde1876} with kind permission of the Hochschul- und Landesbibliothek Fulda, Hesse.}
	\label{fig_transit_instrument}
\end{figure}

The main instrument was a transit telescope with a kinked tube design (focal length 596 mm, aperture 54 mm) which Gerling bought from Ertel \& Sohn (Fig.\ \ref{fig_transit_instrument}). For meridional and prime vertical observations a Kessels box chronometer was used. A meridian stone painted with a scale was placed in about 4 km distance north of the observatory for easy alignment of the telescope.\cite{Gerling1845b} Later in 1862 two further stone marks each painted with a scale were placed in East and West direction of the observatory for use in connection with the prime vertical observations.\cite{Maurutius1862}

Furthermore Gerling's equipment included: a 5 feet Fraunhofer retractor, an English 4 feet refractor on an equatorial mount, a small refractor with larger focal ratio and a prism sextant.\cite{Gerling1843b}. For differential position measurements a ring--micormeter was available.\cite{Westphalen1842} The small house was used for storage and casual observations with the small telescopes.

After finishing the reconstruction of the new institute in fall 1841 Gerling put into service his observatory on 12 October 1841 with an observation of Polaris.\cite{Gerling_316a}
The equipment of the observatory mainly was aimed at the education of young scientists. Gerling and his students did not start a continuous observational program nor did they participate in campaigns. However there exist reports on observations of eclipses, occultations, planets and their moons, asteroids, comets and meteor showers.\cite{Gerling_observations} 
The coordinates for the observatory were determined by Gerling\cite{Gerling1843c} and a small improvement was achieved by
Richard Mauritius, one of Gerling's students:\cite{Maurutius1862b1}

\begin{tabular}{lr}
	longitude & 8\degr\ 46\arcmin\ 24.3\arcsec\ E \\
	latitude  & 50\degr\ 48\arcmin\ 44.1\arcsec\ N \\
	height & 263.7 m
\end{tabular}

A recent GPS measurement resulted in coordinates shifted by about 15\arcsec\ to the West.\cite{Schrimpf2010b1}
The observatory is referred to observatory code 525 at the Minor Planet Center (MPC) of the International Astronomical Union.

Franz Melde, Gerling's successor, mainly used the observatory for time determinations.\cite{Melde1876b1} 
The most well known director of the observatory besides Gerling surely was Alfred Wegener, who had a job as assistant professor at Marburg University from 1909 to 1919. He was not involved in any research at the observatory, though.
The last observations at Gerling Observatory, dealing with variable stars, were reported in the 1930's.

\section{Observations and orbital calculations of asteroids in Marburg}

Gerling mentioned his first attempts of observing asteroids, (6) Hebe and (7) Iris, in a letter to Gauß in fall 1847; but due to bad weather conditions and a deterioration of his eyes Gerling failed.\cite{Gerling_367} 
\emph{In 1849 Gerling's student Wilhelm Klinkerfues carried out the first astrometric observations of an asteroid at Marburg observatory on the 1801 discovered (1) Ceres.\cite{Klinkerfues1850}}

From 1851 to 1852 Eduard Schönfeld started his scientific education at Marburg University and joined Gerling's group. On 27 May 1851 and the following days he observed (14) Irene 8 days after its discovery by John Hind at Greenwich Observatory. His measurements ware published in the \emph{Astronomische Nachrichten}.\cite{Schoenfeld1851} Gerling also mentioned these observations in a letter to Gauß.\cite{Gerling_376}
Surprisingly Schönfeld could find Irene eight days after its discovery at a time, where data about Irene had not yet been published. Presumably, Gerling had received news about Irene by a letter, but unfortunately no note about Irene could be found in the Gerling Archive so far. Besides the data from the observatories in Hamburg, Bonn and Berlin the Marburg observations were among the early observational data from German astronomers.\cite{MPCIrene}
Later in 1852 Schönfeld calculated a small correction for the Marburg Irene data and reported them together with an observation of (6) Hebe.\cite{Schoenfeld1852} With a letter of recommendation from Gerling Schönfeld left Marburg in April 1852 and continued his education under supervision of Friedrich Wilhelm August Argelander at Bonn Observatory.

In fall 1849 Otto Lesser came to Gerling at Marburg University as a student. His first publication -- together with E. Schönfeld -- dealt with the calculation of orbital element of the third comet in 1851.\cite{Schoenfeld1852b} In spring 1852 shortly after the announcement of the discovery of (17) Thetis he observed this asteroid \cite{Gerling_381}
and calculated orbital elements and an ephemeris from observations at Berlin and Hamburg.\cite{Lesser1852} At the beginning of July 1852 he observed (18) Melpomene,\cite{Melpomene} 
which was discovered on 24 June 1852. Lesser left Marburg in fall 1852 and could find an employment with Encke at Berlin Observatory and continued his studies of asteroids. On 14 September 1860 together with Wilhelm Förster he discovered the asteroid (62) Erato. In 1861 he returned back to Gerling for continuing  his studies of the perturbations of the outer planets on the orbits of the asteroids (9) Metis and (32) Pomona.\cite{Lesser1861} He finished his PhD thesis on the perturbations of Jupiter, Saturn and Mars on the orbit of (9) Metis in 1861; a summary was published in the \emph{Astronomische Nachrichten}.\cite{Lesser1863} Lesser left Marburg again and continued his research on asteroid observations and orbital analyses at Altona Observatory.

The published asteroid observations during Gerling's creative period will be appreciated in the next section. No further activity in asteroid research was reported by Gerling's successors.

\subsection{Analysis of the asteroid observations at Marburg}

So far the Marburg (17) Irene observations are the only ones included in the database of the MPC.
To get an idea about the accuracy of the historical asteroid observations in Marburg and to prepare them for insertion into the MPC database the published observational data were compared with orbital calculations using the NASA Horizons On--Line Ephemeris System.\cite{Girogini2017} For all local calculations the software packages Astropy\cite{astropy} and NOVAS\cite{novas} were used.

In a first step the given dates of observations have to be converted to universal time (UT). For the Ceres observations the local mean sidereal time and for all other measurements the local mean solar time 
with an offset of 12 hours was given. For conversion of these dates to UT the geocentric longitude and latitude listed above have been used. 

All asteroid observations at Marburg were carried out extra--meridional using a ring--micrometer by comparing the transits of asteroid and a comparison star passing through a ring in the focal plane of the eyepiece. Using this method it must have been difficult to determine right ascension and declination in one and the same observation. In most cases apparent RA and DEC were listed at different observation times.
Unfortunately no lab notes with further details of these observations could be found in Marburg. As all published observations were made by Gerling's students, they probably took their personal lab notes when they left Marburg. 

For change of equator and equinox to epoch J2000.0 we concatenate observations within a couple of minutes, 23 minutes is the largest time between the observations of the first listed pair of (18) Melpomene measurements. Aberration and nutation 
corrections were applied to the apparent coordinates before they were rotated according to the precession from observational date to J2000.0. Finally coordinates of the asteroids were calculated for the given observational dates and the coordinates of the observatory using the NASA Horizons software. Transformed asteroid coordinates and difference to NASA coordinates are listed in table \ref{T_astr_mr_conv}.

\textbf{(Remove: Obviously there is a systematic error of about 30 \ldots 40\arcsec\ in RA and about half of that in DEC for the Ceres observations.)} 
The errors of \textbf{(Replace: ``all further'' with ``the'')} observations are in the range of 1 \ldots 2\arcsec\ with a few outliers, very well in the range of the accuracy of measurements from other observatories at that time listed in the MPC database. In 1842 Hermann Libert Westphalen, a student in Gerling's group who later worked as a assistant to Bessel at Königsberg, characterized the Marburg ring--micrometer and determined an uncertainty of 3.0655\arcsec,\cite{Westphalen1842b} which matches nicely to the error of the observations.

The asteroid positional data calculated from the observations at Gerling Observatory Marburg have been transferred to the Minor Planet Center (MPC) for insertion into the database.

\begin{landscape}
\begin{table}[!p]
\small\sf\centering
\caption{Asteroid positional data calculated from the observations at Gerling Observatory Marburg. The observations have been converted to J2000 coordinates. The last two columns show the deviations in arcsec from the NASA positional data. For details see text.  \label{T_astr_mr_conv}} \ \\[1em]
\begin{tabular}{llllrr}
\toprule
asteroid & \multicolumn{1}{c}{date} & \multicolumn{1}{c}{RA J2000} & \multicolumn{1}{c}{DEC J2000}  & \multicolumn{1}{c}{$\Delta\text{RA}$}  & \multicolumn{1}{c}{$\Delta\text{DEC}$} \\ 
 &  \multicolumn{1}{c}{(UT)} & \multicolumn{1}{c}{(hms)} & \multicolumn{1}{c}{(dms)}  & \multicolumn{1}{c}{(\arcsec)} & \multicolumn{1}{c}{(\arcsec)} \\ \midrule
(1) Ceres & 1849-06-21 00:14:03.136 & 18 45 57.71 & -27 07 28.67 & -13.981 & -8.918 \\
(1) Ceres & 1849-06-21 00:29:45.556 & 18 45 57.44 & -27 07 22.73 & -8.600 & 0.376 \\
(1) Ceres & 1849-06-22 23:47:23.389 & 18 44  4.82 & -27 16 09.04 & -14.934 & -1.870 \\
(1) Ceres & 1849-06-23 22:13:48.205 & 18 43 12.06 & -27 20 13.42 & -0.279 & -1.846 \\
(1) Ceres & 1849-06-23 22:13:48.205 & 18 43 12.20 & -27 20 16.39 & 1.756 & -4.867 \\
(1) Ceres & 1849-06-23 22:59:52.636 & 18 43 10.11 & -27 20 17.84 & -1.239 & 2.200 \\
(1) Ceres & 1849-06-23 22:59:52.636 & 18 43  9.55 & -27 20 11.96 & -9.726 & 8.178 \\
(14) Irene & 1851-05-28 22:28:24.080 & 16 03 36.78 & -14 04 06.16 & 2.801 & -0.775 \\
(14) Irene & 1851-05-31 22:58:47.180 & 16 00 41.64 & -14 11 07.76 & -9.996 & 2.685 \\
(14) Irene & 1851-06-01 22:48:54.180 & 15 59 46.47 & -14 13 39.00 & 0.731 & -0.614 \\
(14) Irene & 1851-06-02 22:34:05.580 & 15 58 51.62 & -14 16 08.64 & 0.765 & 1.582 \\
(6) Hebe & 1851-07-01 00:07:24.280 & 19 37 32.32 & -07 30 58.96 & -0.619 & -2.808 \\
(18) Melpomene & 1852-07-08 22:58:27.180 & 18 06 45.12 & -09 06 21.83 & 0.304 & -5.014 \\
(18) Melpomene & 1852-07-09 22:41:45.880 & 18 05 48.55 & -09 11 15.61 & 3.935 & 2.127 \\
(18) Melpomene & 1852-07-10 22:41:01.580 & 18 04 51.89 & -09 16 28.51 & 1.886 & 1.816 \\
(18) Melpomene & 1852-07-11 22:22:45.380 & 18 03 57.01 & -09 21 41.84 & 0.781 & 5.548 \\
\bottomrule
\end{tabular}
\end{table}
%
\end{landscape}

\section{Conclusion}

Gerling got in contact with asteroid research early in his career while he was working with Gauß and Harding in Göttingen. During his time in Kassel he took care of orbit and ephemeris calculations. At Marburg University he was participating in improving star maps and after he put into operation his new observatory he supervised his student's observations and perturbation calculations, thus ensuring the continuation in the next generation. 
Gerling started his research on asteroids with an analysis of the Vesta orbit. Shortly before his death he was involved in calculations of the influence of the large planets on the orbits of asteroids just at the threshold of Kirkwood's discovery of the gaps in the distribution of periods of the main belt asteroids.
The asteroids definitely have been a lifelong topic of Gerling's research reflecting the latest developments.

Nowadays the search for solar system objects mainly is done with robotic survey telescopes and automated software pipelines analysing differences of succeeding observations. However, it was the enthusiasm of our predecessors, the commitment to go to the limits of available technique in the 19th century, that laid the path to modern solar system astronomy. 

One of the most successful German groups in searching for new asteroids without a robotic survey programme are the astronomers of the \emph{Frankfurter Physikalische Verein}, Erwin Schwab and Rainer Kling, using the Hans--Ludwig--Neumann Observatory (Code B01) North West of Frankfurt, Hesse; they discovered more than 100 new asteroids in the last 20 years. On 11 February 2008 a new minor planet, 2008 CW16, could be detected\cite{Schwab2017}
and in the next years the orbit was determined with high precision. After proofing that no earlier identifications exist they suggested to name it ``Marburg''. On 12 October 2011 the MPC published this name in the circular \# 76677:

\begin{itemize}
\item 
(256813) Marburg = 2008 CW116\\
Discovered 2008 Feb. 11  by E. Schwab and R. Kling at Taunus.\\

Marburg is a German city, first mentioned in 1138.  A castle and Germany’s oldest gothic church dominate the medieval cityscape.  In 1527 the oldest protestant university in the world was founded, which has operated the Marburg Observatory (later renamed Gerling Observatory) since 1841.
\end{itemize}

Gerling's contribution to the asteroid research is at the same level as that of others of that time and should not be missed. Naming an asteroid after the city where he founded his observatory seems to be an adequate appreciation of Gerling's and his students' research.

\begin{acks}
We like to thank Gesine Brakhage and Dr. Bernd Reifenberg for their very helpful support in searching and reviewing the archive of Christian Ludwig Gerling at the library of the Philipps--Universität Marburg as well as scanning some of the letters for reproduction, Rudi Czech for help in the transliteration of 19th century handwritings and Judith Whittaker--Stemmler for the translation of German quotations. We also acknowledge an anonymous referee for helpful suggestions to strengthen the central idea of the paper.

This research made use of Astropy, a community--developed core Python package for Astronomy (Astropy Collaboration, 2013) and of data and/or services provided by the International Astronomical Union's Minor Planet Center and the Solar System Dynamics Group of the Jet Propulsion Laboratory. 
\end{acks}

\begin{funding}
This research received no specific grant from any funding agency in the public, commercial, or not--for--profit sectors. 
\end{funding}

\begin{sm}

\selectlanguage{german}

\noindent
\textbf{Transliteration of a letter from C.F.\ Gauß to Chr.L.\ Gerling dated 16 April 1814}

In great haste I am sending you here, dearest Gerling, a manuscript from Bessel’s observations of Vesta, which I have just received. You will have to take the geographic location of Königsberg from the Conn. des Tems, for he failed to name the exact position of the new observatory’s site to me, also.
\vspace{1em}

\noindent
$<$List of observational data from 31 January 1814 to 26 February 1814: date and time, RA, DEC$>$
\vspace{1em}

\noindent
May I ask you to disclose the tableau of comparisons of all observations to me first before you derive the opposition from it.

\noindent
Göttingen, 16 April 1814	\hspace{3cm}			Most respectfully yours, CFG

\vspace{1em}

\noindent
Concerning Encke, you will receive the news from your fellow countryman himself.

\noindent
The declinations that are printed in the Göttinger Gelehrten Anzeigen (Academic Journal of Göttingen) were sent in later to B. and G.

\end{sm}

\section{Notes on Contributors}
Julia Remchin studied physics and mathematics for teaching at high schools at Justus--Liebig Universität, Gießen, Germany and at Philipps--Universität Marburg, Germany; finishing in 2017. She joined the group History of Astronomy and Observational Astronomy, Philipps--Universität Marburg, and reviewed and compiled the historical documents about Gerling's asteroid research for her thesis.

Andreas Schrimpf studied physics  at Philipps--Universität Marburg, Germany; 1986 PhD in solid state physics; 1987 --- 1988 fellow of the Deutsche Forschungsgemeinschaft at University of Virginia, USA; 1994 habilitation in solid state physics at Philipps--Universität Marburg; 1995/1996 visiting professorship at Universität Kassel; starting in 1996 research staff member at Physics Department, Philipps--Universität Marburg; starting 2004 head of workgroup on History of Astronomy and Observational Astronomy at Physics Department, Philipps--Universität Marburg; 2015 associate professor at Physics Department, Philipps--Universität Marburg;
publications in solid state physics, history of astronomy; recent projects: history of astronomy, focus on Hesse, Germany; observation and analysis of variable stars, focus on automated analysis of the Sonneberg Plate Archive.



\end{document}